\documentclass[manuscript,screen]{acmart}

\AtBeginDocument{%
  }

\setcopyright{acmlicensed}
\acmJournal{TOIS}
\acmYear{2025} \acmVolume{1} \acmNumber{1} \acmArticle{1}
\acmMonth{1}\acmDOI{10.1145/3728465}

\acmConference[Conference acronym 'XX]{Make sure to enter the correct
  conference title from your rights confirmation emai}{June 03--05,
  2018}{Woodstock, NY}
\acmISBN{978-1-4503-XXXX-X/18/06}

\usepackage{enumitem}
\usepackage{multirow}
\usepackage{bbding}
\usepackage{bm}
\usepackage{makecell}

\begin{document}

\title{User Invariant Preference Learning for Multi-Behavior Recommendation}

\author{Mingshi Yan}
\affiliation{%
  \institution{Tianjin University}
  \city{Tianjin}
  \country{China}
}
\email{neo.ms.yan@gmail.com}

\author{Zhiyong Cheng}
\affiliation{%
  \institution{Hefei University of Technology}
  \city{Hefei}
  \country{China}
}
\email{jason.zy.cheng@gmail.com}

\author{Fan Liu}
\affiliation{%
  \institution{National University of Singapore}
  \city{Singapore}
  \country{Singapore}
}
\email{liufancs@gmail.com}

\author{Yingda Lyu}
\affiliation{%
  \institution{Jilin University}
  \city{Jilin}
  \country{China}
}
\email{ydlv@jiu.edu.cn}

\author{Yahong Han}
\authornote{Y. Han is the corresponding author.}
\affiliation{%
  \institution{Tianjin University}
  \city{Tianjin}
  \country{China}
}
\email{yahong@tju.edu.cn}







\renewcommand{\shortauthors}{Yan et al.}

\begin{abstract}

  In multi-behavior recommendation scenarios, analyzing users' diverse behaviors, such as \textit{click}, \textit{purchase}, and \textit{rating}, enables a more comprehensive understanding of their interests, facilitating personalized and accurate recommendations. A fundamental assumption of multi-behavior recommendation methods is the existence of shared user preferences across behaviors, representing users' intrinsic interests. Based on this assumption, existing approaches aim to integrate information from various behaviors to enrich user representations. However, they often overlook the presence of both commonalities and individualities in users' multi-behavior preferences. These individualities reflect distinct aspects of preferences captured by different behaviors, where certain auxiliary behaviors may introduce noise, hindering the prediction of the target behavior.
  To address this issue, we propose a user invariant preference learning for multi-behavior recommendation (UIPL for short), aiming to capture users' intrinsic interests (referred to as invariant preferences) from multi-behavior interactions to mitigate the introduction of noise. Specifically, UIPL leverages the paradigm of invariant risk minimization to learn invariant preferences. To implement this, we employ a variational autoencoder (VAE) to extract users' invariant preferences, replacing the standard reconstruction loss with an invariant risk minimization constraint. Additionally, we construct distinct environments by combining multi-behavior data to enhance robustness in learning these preferences. Finally, the learned invariant preferences are used to provide recommendations for the target behavior. Extensive experiments on four real-world datasets demonstrate that UIPL significantly outperforms current state-of-the-art methods.

\end{abstract}



\maketitle

\section{Introduction}

In recent years, recommendation systems~\cite{SGL, DiffRec, DDRM} have become increasingly prevalent across industries such as the Internet, e-commerce, and media, highlighting their growing significance. These systems enhance the user experience while driving business value for enterprises by improving the efficiency of information dissemination. By analyzing user preferences, recommendation systems can deliver personalized content and products, thereby boosting user satisfaction and loyalty. Collaborative filtering~\cite{LightGCN, MEGCF, ZhouZY23} (CF) has emerged as the predominant approach in recommender systems, owing to its simplicity and efficacy. Over the past few decades, CF-based models have undergone a transformative journey, progressing from shallow models like matrix factorization~\cite{BPR, BPRH} to more sophisticated architectures, such as deep neural networks~\cite{ZhuLCLZ20, NMTR} and graph neural networks~\cite{MBGCN, MB-CGCN}.

Initially, CF-based methods focus on modeling a single type of historical interaction~\cite{DGCF, ChenXLHL23, WuHWZW23}, such as user purchase records, often encountering data sparsity issues. Fortunately, alternative types of interaction data, such as \textit{click} and \textit{collect} behaviors in e-commerce platforms, are available and contain rich user preferences. Consequently, a growing body of research~\cite{CRGCN, HMG-CR, MB-CGCN} has incorporated these diverse types of interactions into the modeling process to alleviate data sparsity. These methods, which leverage multiple types of interaction data, are known as multi-behavior recommendations (MBR). Within these approaches, the behavior that receives the most attention, such as the \textit{purchase} behavior in e-commerce platforms, is called the target behavior. In contrast, other behaviors, such as \textit{click} and \textit{collect}, are classified as auxiliary behaviors.

The core idea of MBR lies in extracting useful information from auxiliary behaviors to enhance user representations~\cite{CKML, GHCF} and improve the accuracy of target behavior recommendations. Existing MBR methods primarily follow two technical approaches. The first approach focuses on integrating representations from different behaviors~\cite{MBGCN, MATN, EHCF}. These methods typically learn user preferences for each behavior independently and then merge them using specific aggregation strategies. For example, MBGCN~\cite{MBGCN} and s-MBRec~\cite{SMBREC} employ graph convolutional networks (GCN) to model user and item representations for each behavior. MBGCN computes behavior weights based on interaction quantities for weighted fusion, while s-MBRec applies linear and non-linear fusion techniques for users and items, respectively. The second approach focuses on exploring relationships between behaviors by jointly modeling multi-behavior data~\cite{CRGCN, MB-CGCN, PKEF}. Methods like NMTR~\cite{NMTR} calculate similarity scores between users and items in previous behaviors to influence subsequent interactions, while CRGCN~\cite{CRGCN} and MB-CGCN~\cite{MB-CGCN} use embeddings from earlier behaviors as initial values for later ones to establish inter-behavior relationships. As MBR research develops, the challenge of noise in representations derived from auxiliary behaviors has gained attention, as it can negatively affect the predictive accuracy of target behaviors. Recent methods address this issue by mitigating the impact of noise. For instance, PKEF~\cite{PKEF} achieves information decoupling through the projection mechanism, and MBSSL~\cite{MBSSL} employs contrastive learning at inter-behavior and intra-behavior levels to enhance model robustness.

\begin{figure}[hbt]
    \centering
    \includegraphics[width=\linewidth]{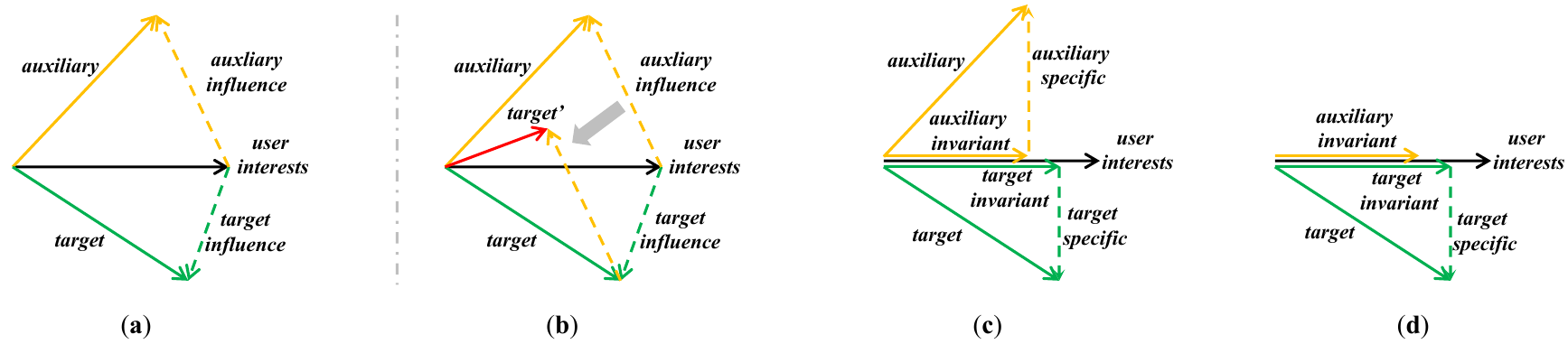}
    \caption{A toy example (provided to illustrate the existence of user-invariant preferences and the noise introduced in modeling auxiliary behavior information).}
    \Description{}
    \label{fig:toys}
\end{figure}

While some existing works recognize the presence of noise, they lack a comprehensive analysis of its sources, leading to certain limitations. Multi-behavior recommendation methods rest on the assumption that user preferences across different behaviors share similarities, reflecting users' intrinsic interests. Consequently, these methods focus on integrating information from various behaviors to enrich user representations. However, they overlook the fact that preferences learned from auxiliary behaviors primarily reflect interests specific to those behaviors, which may include components irrelevant to the target behavior. These irrelevant components primarily arise from external influences, such as seasons, social environments, and emotional states. They are specific to the current behavior and can be regarded as noise in the context of other behaviors. As illustrated in Fig.~\ref{fig:toys}(a), the solid black vector represents user interests, the solid orange vector denotes preferences within auxiliary behaviors, and the dashed orange vector represents the influencing factors driving auxiliary behaviors. These factors shift user interests, leading to interactions in auxiliary behaviors. Similarly, the green vector represents user preferences for the target behavior. Incorporating these auxiliary influences into the target behavior, as shown by the solid red vector in Fig.~\ref{fig:toys}(b), will introduce biases that degrade the accuracy of recommendations, effectively acting as noise. To address this issue, we extract users' intrinsic interests from multi-behavior data. As shown in Fig.~\ref{fig:toys}(c), user preferences within a given behavior can be decomposed into two parts: invariant preferences (representing user intrinsic interests) and behavior-specific preferences (driven by external factors). The projection of user preferences from auxiliary behaviors is consistent with that from target behaviors to user interests, indicating that the user's invariant preferences remain consistent. Therefore, as illustrated in Fig.~\ref{fig:toys}(d), improving target behavior recommendation performance and preventing noise introduction can be achieved by incorporating only the invariant preferences extracted from auxiliary behaviors. This shifts the challenge to effectively extracting users' invariant preferences from multi-behavior data.

Based on the discussion above, we propose a user invariant preference learning (UIPL) method for multi-behavior recommendation, aiming at learning users' invariant preferences from various behaviors to enhance the recommendation performance of the target behavior. UIPL leverages the invariant risk minimization (IRM) paradigm to achieve this goal, which emphasizes learning stable representations of shared features across multiple environments. 
In the IRM paradigm, environments refer to different data distribution spaces, and multi-behavior data aligns well with this concept. Therefore, each behavior can be treated as a separate environment, enabling IRM to effectively capture the shared user preferences across different behaviors, namely the invariant preferences. In theory, a greater number of environments can provide stronger constraints~\cite{IRMv1, IRL}. To enhance the learning of invariance, we expand the set of environments using existing multi-behavior data (refer to Section~\ref{erl} for more details).
To accomplish this, we design a variational autoencoder (VAE) network to generate invariant preferences. The encoder extracts critical features from user embeddings and maps them to a latent space, leveraging its feature extraction and data compression capabilities. The decoder then maps the latent representation into the invariant preference space to reconstruct the user’s invariant preference. Notably, the decoder's goal is not to reconstruct the original data but to generate the invariant preference. Hence, we employ the IRM loss instead of a traditional reconstruction loss. To better constrain the generation of the invariant preference, we also introduce contrastive and orthogonal loss to assist in model training. Since data from multiple environments contribute to learning shared feature invariance via IRM, we combine multi-behavior data to create multiple environments for representation learning. This approach enables the extraction of multiple user representations, facilitating IRM in learning the invariant preferences. Finally, we aggregate the learned user invariant preferences from different environments to provide recommendations for the target behavior. Extensive experiments on four datasets demonstrate the effectiveness of our approach. Compared to state-of-the-art models, our method consistently achieves substantial performance improvements across all datasets.

In summary, our work presents the following key contributions:
\begin{itemize}[leftmargin=*]
    \item We analyze the sources of noise introduced in multi-behavior modeling, and propose a novel approach to address the noise issue by learning user invariant preferences. As far as we know, this is the first work to model invariant preferences in the multi-behavior recommendation.

    \item We introduce a method for learning invariant preferences based on the IRM paradigm, which incorporates an autoencoder network to generate invariant preferences. Additionally, we leverage multi-behavior data to construct multiple environments, facilitating the implementation of invariant risk minimization.

    \item Extensive experiments on four real-world datasets demonstrate that our method outperforms state-of-the-art models across all datasets. Ablation studies further validate the rationale behind our model design. 
    Furthermore, our code is available on GitHub~\footnote{https://github.com/MingshiYan/UIPL} for research.
\end{itemize}
\section{Related Work}  \label{Related Work}
\subsection{Multi-behavior Recommendation}
Multi-behavior recommendation has achieved significant advancements due to its effectiveness in mitigating data sparsity~\cite{CIGCN, CKML, RCL, HPMR}. At its core, this approach enriches user representations by extracting preferences from auxiliary behaviors, thereby enhancing its recommendation capacity for the target behavior.

Existing multi-behavior recommendation methods mainly follow two different technical approaches. One approach is to learn user preferences independently from each behavior and subsequently aggregate them to form the final user representation. Early methods, exemplified by BPRH~\cite{BPRH}, quantify relevance by assessing user-item co-occurrences across different behaviors to assign weights for representation aggregation. Subsequent methods such as MBGCN~\cite{MBGCN} and GNMR~\cite{GNMR} employ weighted aggregation, differing in the way weights are assigned. These methods focus on accurately capturing user preferences from each behavior using effective approaches. The other approach delves into the interconnections between different behaviors to capture their mutual influences and iteratively refine user preferences. For example, in sequential modeling approaches like NMTR~\cite{NMTR}, CRGCN~\cite{CRGCN}, and MB-CGCN~\cite{MB-CGCN}, NMTR propagates the prediction scores of the current behavior to influence the predictions for subsequent behaviors; CRGCN and MB-CGCN employ embedding propagation to transfer information between multiple behaviors. The core of these methods is to establish the influences between different behaviors, thereby indirectly improving the accuracy of user preferences.

Recent studies have started to recognize the issue of noise in multi-behavior modeling and have proposed solutions. For example, PKEF~\cite{PKEF} employs a projection mechanism in the multi-task module to decouple complex representations and eliminate noise. MBSSL~\cite{MBSSL} addresses the problem of interaction noise by incorporating contrastive learning at both intra- and inter-behavioral levels. BCIPM~\cite{BCIPM} leverages auxiliary behaviors to train a specialized network for more accurate extraction of information from the target behavior, thereby avoiding noise introduction.

In contrast, our work delves deeper into analyzing the root causes of noise and introduces a novel approach for modeling user invariant preferences, offering a more effective solution to the noise problem.

\subsection{Invariant Representation Learning in Recommendation}
Invariant representation learning (IRL)~\cite{HRM, IRMv1, ZhangSFWW0023, DIRL} endeavors to acquire a data representation that retains stability and invariance across diverse environments. This stability facilitates enhanced generalization capabilities of models. IRL has been widely applied in fields such as computer vision, natural language processing, and speech recognition.

In recent years, there has been a growing focus on IRL within recommender systems~\cite{KHRM, IRL, KrishnanDBYS20}. In the traditional recommendation task, IRL has been utilized to mitigate biases present in observed data. For example, InvPref~\cite{InvPref} employs the expectation-maximization (EM) algorithm to partition and update virtual environments while using adversarial learning to learn invariant representations. In the task of multimodal recommendation, researchers adopt invariant risk minimization to learn invariant representations from multimodal data. For instance, InvRL~\cite{InvRL} integrates two iterative modules to cluster user-item interactions into heterogeneous environments and acquire cross-environment invariant representations. Similarly, PaInvRL~\cite{PaInvRL} proposes a method to balance empirical risk loss and invariant risk loss, striving for Pareto optimal solutions. In cross-domain recommendation tasks, learning invariant preferences from different environments are explored to enhance the model's domain-generalization recommendation capability. For example, Grace~\cite{Grace} introduces an adversarial learning approach to discern invariant preferences from domain-specific preferences.

Our work presents a novel model designed to learn invariant representations in multi-behavior scenarios, marking the first attempt to model such representations in multi-behavior recommendation systems. We integrate VAE to generate invariant representations within the IRM framework.
\section{Preliminary} \label{Preliminary}

Before delving into the details of the model, we first revisit our motivation. Our approach is driven by the observation that user invariant preferences, which are consistent across different behaviors, can be leveraged to enhance user representations. By extracting these invariant preferences from user embeddings across various behaviors and aggregating them, we can improve user representations while minimizing the introduction of noise. To achieve this, we introduce the IRM paradigm, which enforces invariance across different environments by applying invariant constraints. Ultimately, we combine the learned invariant preferences to enhance the recommendation performance for the target behavior.

In this section, we first formulate the problem of multi-behavior recommendation. Next, we provide a brief review of the IRM paradigm. Finally, we explain how to effectively apply IRM in multi-behavior scenarios to learn user invariant preferences.

\subsection{Problem Formulation} \label{pf}
We define the user set $\mathcal{U}$ and the item set $\mathcal{I}$, where $u$ and $i$ represent the $u$-th user and the $i$-th item, respectively. $\boldsymbol{P} \in \mathbb{R}^{d \times U}$ and $\boldsymbol{Q} \in \mathbb{R}^{d \times I}$ are the initialization embedding matrices for users and items, respectively. Here, $U$ and $I$ represent the number of users and items, and $d$ is the embedding size. By indexing with $u$ and $i$, we retrieve the corresponding column vectors from $\boldsymbol{P}$ and $\boldsymbol{Q}$ as the representations $\boldsymbol{p}_u$ and $\boldsymbol{q}_i$ for the user $u$ and item $i$, respectively. The set $\mathfrak {R} = \{\mathcal {R}_1, \cdots, \mathcal {R}_k, \cdots, \mathcal {R}_K \}$ signifies a collection of multi-behavior user-item interaction matrices, where $k$ denotes the $k$-th behavior. $\mathcal R_k = \{(u, i) | u \in \mathcal{U} \wedge i \in \mathcal{I} \wedge r_{ui}=1 \}$, where $r_{ui}=1$ indicates an observed interaction between a user and an item (otherwise $r_{ui}=0$). 

Multi-behavior recommendation aims to learn a model $\Gamma(u, i | \Theta)$ that predicts the likelihood of user $u$ interacting with item $i$ in the target behavior. $\Theta$ denotes all the parameters of the model. The formal definition of the optimization objective is as follows:
\begin{equation}
  \label{eq:obj}
   \underset{\Theta}{\arg \min} \mathcal{L}(\Gamma(u, i | \Theta)|\mathfrak R),
\end{equation}
where $\mathcal{L}(\cdot)$ represents an objective function, and we omit the regularization term.

\subsection{Brief of IRM}
Invariant risk minimization (IRM) is a machine learning paradigm~\cite{IRMv1, BIRM, PAIR} aimed at learning representations that exhibit invariance to shared features by minimizing the differences in risk expectations across different environments. At its core, IRM emphasizes the stability of representations regarding shared features or characteristics present in multiple environments. The specific definition is as follows:

\begin{equation}
  \label{eq:ood}
  \begin{aligned}
       \min \limits_{\substack{\Phi: \mathcal{X} \to \mathcal{H} \\ \omega: \mathcal{H} \to \mathcal{Y}}} \quad \sum_{e \in \mathcal{E}} R^e(\omega \circ \Phi), \quad \quad \quad \\
    s.t. \quad \omega \in \underset{\bar{\omega}: \mathcal{H} \to \mathcal{Y}}{\arg \min} R^e(\bar{\omega} \circ \Phi), \ \forall e \in \mathcal{E} ,
  \end{aligned}
\end{equation}
where $\mathcal{X} \to \mathcal{H} (\mathcal{H} \to \mathcal{Y})$ is the functional mapping from the set $\mathcal{X}$ to set $\mathcal{H}$ (set $\mathcal{H}$ to set $\mathcal{Y}$), $R^e$ represents the empirical risk function in environment $e$, $\mathcal{E}$ denotes the set of training environments, and $\circ$ is defined as a multiplicative mapping. This formula seeks to learn an invariant preference $\Phi$ such that the classifier $\omega$ achieves optimal performance across all environments in $\mathcal{E}$. 

Eq.~\ref{eq:ood} describes a challenging bi-level optimization problem, which IRMv1~\cite{IRMv1} addresses through an approximation method:
\begin{equation}
    \label{eq:IRMv1}
    \underset{\Phi}{\min} \sum_{e \in \mathcal{E}} R^e(\Phi) + \lambda ||\nabla_{\omega} R^e(\omega \circ \Phi) ||^2.
\end{equation}

IRMv1 utilizes a regularization term to constrain empirical risk minimization. $R^e(\Phi)$ represents the prediction accuracy in a given training environment $e$. The term $\nabla_{\omega} R^e(\omega \circ \Phi)$ signifies the invariance within the training environment $e$, suggesting that when $\omega$ is optimal in environment $e$, the derivative of $R^e(\omega \circ \Phi)$ approaches zero. $\lambda$ is a hyperparameter used to measure the degree of constraint on invariance. In summary, IRMv1 ensures that features learned in each environment remain invariant across all environments by minimizing the gradient norm.

IRM can be effectively applied to multi-behavior recommendation tasks. As the introduction highlights, a user's inherent interests, shaped by various external factors, manifest as diverse behavioral preferences. These preferences share commonalities across different behaviors, which align with the concept of invariant features in IRM. Therefore, the IRM paradigm can be leveraged to learn shared user preferences in multi-behavior scenarios, namely the user invariant preferences.

Next, we will elaborate on how to apply the IRM paradigm to learn user invariant preferences in a multi-behavior recommendation scenario.

\subsection{IRM for Multi-behavior Recommendation} \label{IPL}
\begin{figure}[hbt]
    \centering
    \includegraphics[width=\linewidth]{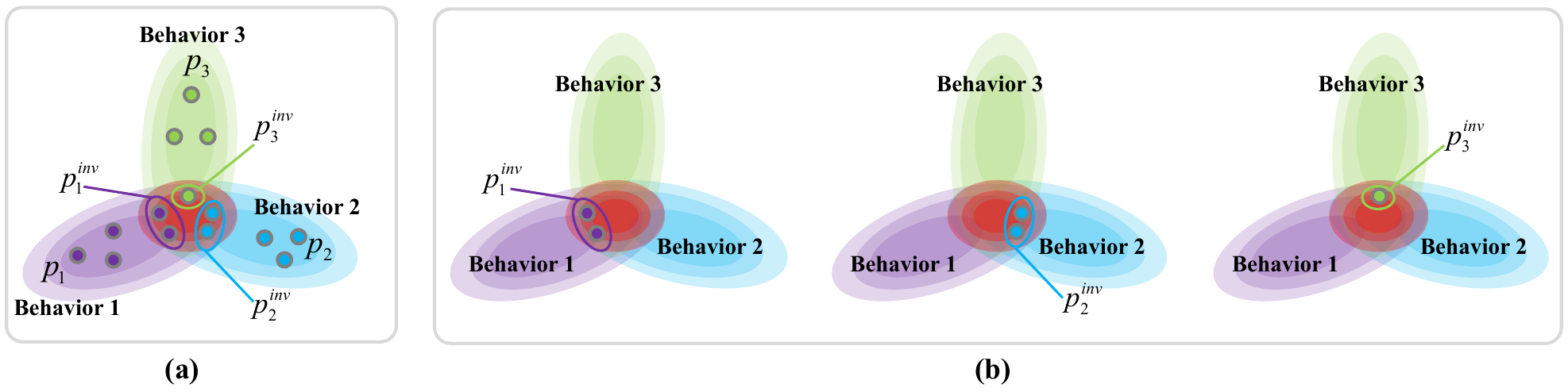}
    \caption{An explanation of IRM for learning invariant preferences ($p_1^{inv}$, $p_2^{inv}$, $p_3^{inv}$ represent the invariant preferences learned from the corresponding behaviors).}
    \Description{}
    \label{fig:supp}
\end{figure}
Please note that our goal is to learn users' invariant preferences. In a multi-behavior scenario, each behavior corresponds to an environment in IRM.  As shown in Fig.~\ref{fig:supp} (a), for a given user, each color represents a behavior space (i.e., environment) containing various preferences of the user. The red area indicates the intersection of different environment spaces. Due to the correlations between behaviors, user preferences falling within the red area are shared across multiple behaviors (i.e., the invariant preferences mentioned earlier). It is precisely due to the existence of this shared nature that the invariant preferences learned from one behavior are also applicable to other behaviors~\cite{PaInvRL}. This allows us to learn the invariant preferences via the IRM paradigm. As illustrated in Fig.~\ref{fig:supp} (b), under the IRM paradigm, we learn invariant preferences from a given behavior (e.g., $p_1^{inv}$ in behavior 1), and use the preference for predicting other behaviors as a constraint term for invariant representation (e.g., predicting behavior 2 and behavior 3 using $p_1^{inv}$), thereby achieving the learning of invariant preferences for each behavior. Hence, Eq.~\ref{eq:IRMv1} can be rewritten as follows:

\begin{equation}
    \label{eq:mb_irm_orig}
    \underset{\Phi^m}{\min}  \sum_{m, n \in \mathcal{E} \wedge n \neq m} R^m(\Phi^m) + \lambda \cdot R^n(\Phi^m),
\end{equation}
where $R^m(\cdot)$ and $R^n(\cdot)$ represent the empirical risk functions for environments $m$ and $n$, respectively. $\Phi^m$ represents the user preferences learned from environment $m$, and $\mathcal{E}$ denotes the set of training environments ($\mathcal{E}$ can be viewed as the set of behaviors here, which will be augmented in Section ~\ref{erl}). Eq.~\ref{eq:mb_irm_orig} applies the empirical risk function to impose the invariance constraint, minimizing the empirical risk in other environments ($R^n$) to ensure the invariance of $\Phi^m$ (learned from environment $m$) across these environments.

In recommendation scenarios, it is common to guide the model to learn user representations by predicting interactions between users and items. Hence, we treat the empirical risk minimization problem as a binary classification task, aiming to predict whether there is an edge connection between nodes in a bipartite graph of user-item interactions. Specifically, we predict whether the element (denoted as $r_{ui}$) of the user-item interaction matrix $\mathcal{R}_m$ in environment $m$ is 1 or 0. Therefore, $\Phi^m$ can be rewritten as:
\begin{equation}
    \label{eq:emr}
    \Phi^m =  f(\phi_u^m, \psi_i | r_{ui}^m),
\end{equation}
where $f(\cdot)$ is an edge connection prediction function, typically using methods such as calculating correlations. $\phi_u^m$ represents the user $u$’s preferences learned from environment $m$, and $\psi_i$ is the item's representation, $r_{ui}^m \in \{0, 1\}$ denotes the ground truth value associated with the interaction matrix $\mathcal {R}_m$, indicating whether the interaction between user $u$ and item $i$ is observed. 
Please note that $\psi_i$ remains unchanged regardless of variations in behavior, which is intuitive since item features are constant. Therefore, in the IRM process, only $\phi_u^m$ is constrained to facilitate the learning of invariant preferences. 
Eq.~\ref{eq:mb_irm_orig} can be rewritten as:
\begin{equation}
    \label{eq:mb_irm_midd}
    \min  \sum_{m, n \in \mathcal{E} \wedge n \neq m} R^m(f(\phi_u^m, \psi_i | r_{ui}^m)) + \lambda \cdot R^n(f(\phi_u^m, \psi_i | r_{ui}^n)),
\end{equation}
where $r_{ui}^m$ in the first term represents the optimization in the $m$-th environment, and $r_{ui}^n$ in the second term represents the optimization in the $n$-th environment based on the user preferences learned from the $m$-th environment (i.e., implementing  the invariant constraint).

Observing Eq.~\ref{eq:mb_irm_midd}, we notice that the sole distinction between the two risk function terms lies in the terms $r_{ui}^m$ and $r_{ui}^n$ ($R^m(\cdot)$ and $R^n(\cdot)$ can utilize the same empirical risk function). The hyperparameter $\lambda$ is used to measure the importance of the invariant constraint. In theory, a larger value of $\lambda$ implies a stronger invariance constraint. However, in recommendation tasks, the importance of user preferences learned from a specific behavior is much higher within that behavior than across other behaviors. Therefore, the value of $\lambda$ should not exceed 1. To simplify the model, we fix $\lambda=1$. Consequently, Eq.~\ref{eq:mb_irm_midd} can be simplified as follows:
\begin{equation}
    \label{eq:mb_irm_fin}
    \min  \sum_{n \in \mathcal{E}} \sum_{m \in \mathcal{E}} R(f(\phi_u^m, \psi_i | r_{ui}^n)),
\end{equation}
when $m = n$, $R(\cdot)$ represents the empirical risk minimization for $\phi_u^m$, and when $m \neq n$, $R(\cdot)$ denotes the invariant risk constraint on $\phi_u^m$. Minimizing Eq.~\ref{eq:mb_irm_fin} enables the optimization of $\phi_u^m$, representing the invariant preferences extracted from environment $m$.

Thus far, we have derived how to utilize IRM to optimize invariant preferences in multi-behavior scenarios. 
To implement Eq.~\ref{eq:mb_irm_fin}, we need to design a model for learning the item representation $\psi_i$ and the user invariant preference $\phi_u^m$.
Next, we will dive into the implementation details of our model and explain how the learned invariant preferences are applied to the target behavior recommendation task.
\section{Methodology} \label{Methodology}

\begin{figure}[t]
    \centering
    \includegraphics[width=\linewidth]{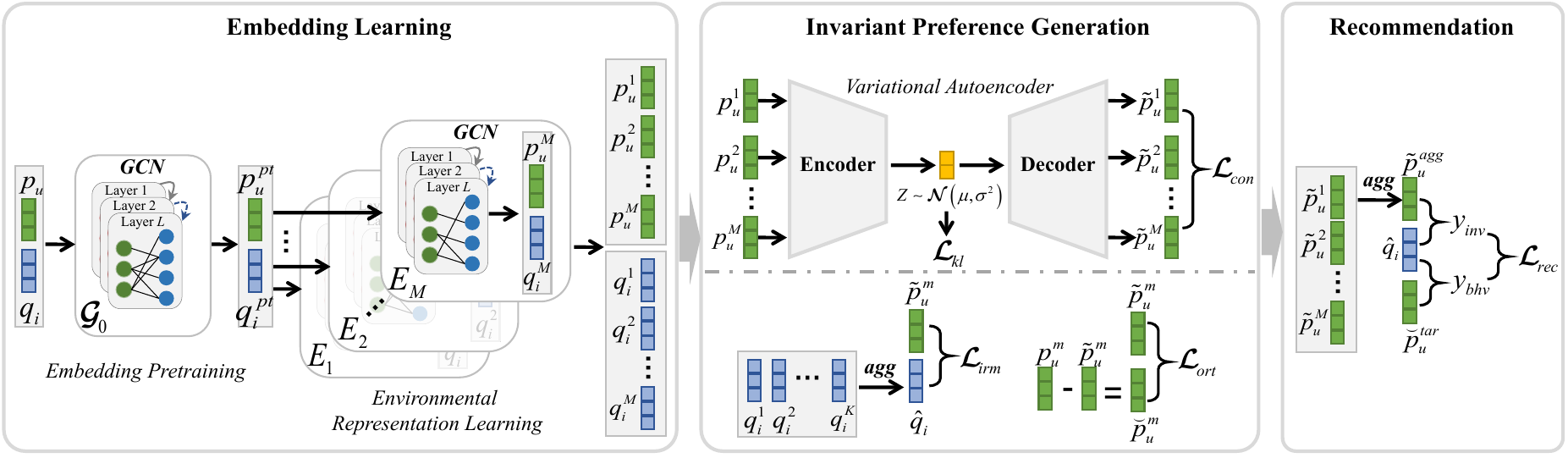}
    \caption{Overview of UIPL ($\{ \boldsymbol{q}_i^1, \boldsymbol{q}_i^2, \cdots, \boldsymbol{q}_i^K \}$ represent the item representations learned from environments constructed independently by each behavior; $\boldsymbol{p}_u^m$ and $\tilde{\boldsymbol{p}}_u^m$ denote the $m$-th input and output of the encoder-decoder model, respectively).}
    \label{fig:global}
    \Description{}
\end{figure}

After the above derivation, we have gained an understanding of how to utilize IRM to model invariant preferences in a multi-behavior scenario. In this section, we will introduce our method, UIPL, which implements the learning of invariant preferences based on the aforementioned derivations and applies them to the target behavior recommendation task. Fig.~\ref{fig:global} illustrates the overall structure of our model: (1) \textbf{Embedding Learning} is responsible for learning user and item embeddings from multi-behavior interactions, which are then used by downstream modules. (2) \textbf{Invariant Preference Generation} aims to generate invariant preferences by leveraging user representations learned across multiple environments. (3) \textbf{Recommendation} is utilized to integrate the learned user invariant preferences for making predictions in the target behavior. 

\subsection{Embedding Learning}
The embedding learning module consists of two components: embedding pretraining and environmental representation learning. The embedding pretraining part focuses on learning coarse-grained representations from interaction records of all behaviors, which helps improve the quality of user representations in the subsequent environmental representation learning phase. In this phase, different combinations of multi-behavior data are used to construct multiple environments, enabling the learning of user representations in diverse contexts. This approach supports the IRM process, facilitating a more effective learning of invariant preferences.

\textbf{\textit{Embedding Pretraining}}.
Since user and item embeddings are typically initialized randomly (by uniform or normal distributions), they may lack meaningful information and are prone to be influenced by variations in data distribution when utilized in a shared manner~\cite{BCIPM}. To address this, we first model all interactions simultaneously to learn a coarse-grained representation, which serves as input for the environmental representation learning phase. Specifically, we construct a new interaction matrix, $\mathcal{R}_0$, by combining user-item interaction matrices from all behaviors ($\mathcal{R}_0 = \mathcal{R}_1 \cup \mathcal{R}_2 \cup \cdot \cdot \cdot \cup \mathcal{R}_M$, where "$\cup$" represents the union operation). We then apply graph convolution operations to the interaction graph constructed from this matrix. For simplicity, we use LightGCN~\cite{LightGCN} for this task. The process is as follows:
\begin{equation}
  \label{eq:gcn}
  \begin{aligned}
    \boldsymbol p_{u}^{(l+1)} &= \sum_{i \in N_{u}} \frac{1}
    {\sqrt{\left\lvert N_{u} \right\rvert} \sqrt{\left\lvert N_{i} \right\rvert}} \boldsymbol q_{i}^{(l)}, \\ 
    \boldsymbol q_{i}^{(l+1)} &= \sum_{u \in N_{i}} \frac{1}
    {\sqrt{\left\lvert N_{i} \right\rvert} \sqrt{\left\lvert N_{u} \right\rvert}} \boldsymbol p_{u}^{(l)}, 
  \end{aligned}
\end{equation}
where $N_{u}$ ($N_i$) represents the items (users) interacted with user $u$ (item $i$) in the interaction matrix $\mathcal{R}_0$. When $l=0$, $\boldsymbol p_{u}^{(0)}$ and $\boldsymbol q_{i}^{(0)}$ are the initial input embeddings (i.e., the initialized embeddings $\boldsymbol p_u$ and $\boldsymbol q_i$ in Section~\ref{pf}). Next, aggregate the embeddings at each layer to serve as representations for users and items:
\begin{equation}
  \label{eq:norm}
  \begin{aligned}
    \boldsymbol{p}^{pt}_{u} = \sum_{l=0}^{L} \boldsymbol{p}_{u}^{(l)}, \quad
    \boldsymbol{q}^{pt}_{i} = \sum_{l=0}^{L} \boldsymbol{q}_{i}^{(l)},
  \end{aligned}
\end{equation}
where $L$ denotes the number of GCN layers, and $\boldsymbol{p}^{pt}_{u}$ and $\boldsymbol{q}^{pt}_{i}$ are taken as the input to the environmental representation learning module.

\textbf{\textit{Environmental Representation Learning}}. \label{erl}
As discussed in Section~\ref{IPL}, the IRM framework ensures that learned user representations can accurately predict across various environments, thus enforcing invariant constraints. To improve the ability to constrain invariant preferences, it is crucial to increase the diversity of the environments~\cite{IRL, IRMv1}. For this, we construct multiple environments by combining the multi-behavior interaction data. Specifically, we generate all possible combinations of elements from the set $\mathfrak {R}$ (defined in Section~\ref{pf}). For instance, given a set of interaction matrices with three behaviors \{$\mathcal{R}_0, \mathcal{R}_1, \mathcal{R}_2$\}, we generate the environment set $\mathcal{E}=\{\mathcal{R}_0, \mathcal{R}_1, \mathcal{R}_2, \mathcal{R}_0 \cup \mathcal{R}_1, \mathcal{R}_0 \cup \mathcal{R}_2, \mathcal{R}_1 \cup \mathcal{R}_2, \mathcal{R}_0 \cup \mathcal{R}_1 \cup \mathcal{R}_2 \}$ ("$\cup$" denotes the union operation), which is used in Eq.~\ref{eq:mb_irm_fin}. This approach is effective because combining different interaction matrices is equivalent to merging their noise and altering data distributions (see Fig.~\ref{fig:toys}). Despite these variations, invariant preferences can still be extracted, and these diverse environments also serve as constraints, aiding the learning of invariant preferences across other environments.

Suppose there are $M$ elements in the environment set $\mathcal{E}$. We utilize LightGCN to learn user and item representations in each environment. Based on Eq.~\ref{eq:gcn} and ~\ref{eq:norm}, taking $\boldsymbol{p}^{pt}_{u}$ and $\boldsymbol{q}^{pt}_{i}$ as inputs for each environment, we obtain sets $\{ \boldsymbol{p}_u^1, \boldsymbol{p}_u^2, \cdots, \boldsymbol{p}_u^M \}$ and $\{ \boldsymbol{q}_i^1, \boldsymbol{q}_i^2, \cdots, \boldsymbol{q}_i^M \}$, which represent the learned user and item embeddings, respectively, across the $M$ environments.

\subsection{Invariant Preference Generation}
The generation of invariant preferences is implemented by a VAE network. In this network, the encoder extracts features from the input embeddings and compresses them into a latent space, while the decoder reconstructs user representations from this latent space. The core of our approach is the use of the IRM constraint loss (i.e., Eq.~\ref{eq:mb_irm_fin}), rather than the standard reconstruction loss typically employed in VAEs~\cite{VAE}. This ensures the generated user representations capture the user's invariant preferences across different behaviors. Additionally, to further enhance the learning of these invariant preferences, we incorporate orthogonal and contrastive loss as supplementary components to optimize the model more effectively.

\textbf{\textit{Encoder}}.
The encoder employs a two-layer neural network to extract latent features from the input user embeddings. Given the user embedding from the $m$-th environment, the computational process in the encoder is as follows:
\begin{equation}
  \label{eq:encoder}
   \boldsymbol{h} = \text{relu} (\mathcal{W}_1 \cdot \boldsymbol{p}_u^m + \boldsymbol{b}_1),
\end{equation}
\begin{equation}
  \label{eq:mu}
   \boldsymbol{\mu} = \mathcal{W}_2 \cdot \boldsymbol{h} + \boldsymbol{b}_2,
\end{equation}
\begin{equation}
  \label{eq:sigma}
   \boldsymbol{\sigma}^2 = \mathcal{W}_3 \cdot \boldsymbol{h} + \boldsymbol{b}_3,
\end{equation}
where $\text{relu}(\cdot)$ denotes an activation function, $\mathcal{W}_1 \in \mathbb{R}^{\frac{d}{2} \times d}, \mathcal{W}_2, \mathcal{W}_3 \in \mathbb {R}^{\frac{d}{4} \times \frac{d}{2}}$, and $\boldsymbol{b}_1 \in \mathbb{R}^{\frac{d}{2}}, \boldsymbol{b}_2, \boldsymbol{b}_3 \in \mathbb{R}^{\frac{d}{4}}$ denote the parameters of the network, $\boldsymbol{\mu}$ and $\boldsymbol{\sigma}^2$ represent the mean and variance, respectively, of the Gaussian distribution for the latent variable output by the encoder, following the standard approach of VAEs~\cite{VAE}.

\textbf{\textit{Decoder}}.
Similar to the encoder, the decoder also utilizes a two-layer network for implementation:
\begin{equation}
  \label{eq:decoder}
   \boldsymbol{\tilde{p}}_u^m = \mathcal{W}_5(\text{relu} (\mathcal{W}_4 \cdot \boldsymbol{z} + \boldsymbol{b}_4)) + \boldsymbol{b}_5,
\end{equation}
\begin{equation}
  \label{eq:z}
   \boldsymbol{z} = \boldsymbol{\mu} + \boldsymbol{\sigma} \odot \boldsymbol{\epsilon}, \  \boldsymbol{\epsilon} \sim \mathcal{N}(0, \boldsymbol{I}),
\end{equation}
where $\mathcal{W}_4 \in \mathbb{R}^{\frac{d}{2} \times \frac{d}{4}}, \mathcal{W}_5 \in \mathbb {R}^{d \times \frac{d}{2}}$, and $\boldsymbol{b}_4 \in \mathbb{R}^{\frac{d}{2}}, \boldsymbol{b}_5 \in \mathbb{R}^{d}$ denote the parameters of the network, $\boldsymbol{\epsilon}$ is randomly sampled from the standard normal distribution, and $\boldsymbol{\tilde{p}}_u^m$ represents the output of the decoder, indicating the invariant preferences extracted from user representations in the $m$-th environment. Eq.~\ref{eq:z} illustrates the re-parameterization trick~\cite{VAE}, which resolves the non-differentiability issue associated with the sampling operation.

\textbf{\textit{IRM Loss}}.
To achieve Eq.~\ref{eq:mb_irm_fin}, we need to obtain the invariant user representations, denoted as $\phi_u^m$, and item representations, denoted as $\psi_i$. Through encoding and decoding operations, we derive the user invariant preferences set $\{ \boldsymbol{\tilde{p}}_u^1, \boldsymbol{\tilde{p}}_u^2, \cdots, \boldsymbol{\tilde{p}}_u^M \}$ across all environments. For items, since their features remain constant, we can share item representations across all environments. Thus, we aggregate the item embeddings obtained during the embedding learning process to form the final item representations. To mitigate potential redundancy in information within environments constructed from combinations of multiple behaviors, we select item embeddings learned from environments originating from a single behavior (i.e., each behavior represents an environment) to form the set $\{ \boldsymbol{q}_i^1, \boldsymbol{q}_i^2, \cdots, \boldsymbol{q}_i^K \}$, and aggregate these embeddings as the final item representation:
\begin{equation}
  \label{eq:sum_q}
   \boldsymbol{\hat{q}}_i = \sum_{k=1}^K \boldsymbol{q}_i^k.
\end{equation}

As discussed earlier, user invariant preferences can be determined indirectly by predicting edges through the calculation of user-item correlations, as shown in Eq.~\ref{eq:emr}. Consequently, Eq.~\ref{eq:mb_irm_fin} can be implemented as follows:
\begin{equation}
    \label{eq:irm_loss}
    \mathcal{L}_{irm} = -\sum_{m \in \mathcal{E}} \sum_{\substack{n \in \mathcal{E} \\ u \in \mathcal{U}, i \in \mathcal{I}}} r_{ui}^n \log ( \sigma((\boldsymbol{\tilde{p}}_u^m)^{\top} \cdot \boldsymbol{\hat{q}}_i)) + (1 - r_{ui}^n) \log (1 - \sigma((\boldsymbol{\tilde{p}}_u^m)^{\top} \cdot \boldsymbol{\hat{q}}_i)),
\end{equation}
where $r_{ui}^n \in \{0, 1 \}$ represents the element in the user-item interaction matrix for the $n$-th environment, and $\sigma(\cdot)$ denotes the sigmoid function. For clarity and convenience of description, we use $\boldsymbol{\tilde{p}}_u^m$ ($\boldsymbol{\tilde{p}}_u^m \in \{ \boldsymbol{\tilde{p}}_u^1, \boldsymbol{\tilde{p}}_u^2, \cdots, \boldsymbol{\tilde{p}}_u^M \}$) and $\boldsymbol{\hat{q}}_i$ instead of $\phi_u^m$ and $\psi_i$, respectively.

\textbf{\textit{Orthogonal Loss}}.
In the previous, we discussed the decomposition of user preferences into invariant preferences and behavior-specific preferences (see Fig~\ref{fig:toys}(c)). Through encoding and decoding operations, we extract the user's invariant preferences, enabling us to compute the behavior-specific preferences by calculating the difference. For a given user preference, we expect the decomposed invariant and behavior-specific preferences to be uncorrelated for maximizing information and ensuring stability. To enforce this constraint, we introduce orthogonal loss, which is implemented by minimizing the correlation between invariant preferences and behavior-specific preferences:
\begin{equation}
    \label{eq:ort_loss}
    \mathcal{L}_{ort} = \frac{1}{|\mathcal{U}|} \frac{1}{|\mathcal{E}|} \sum_{u \in \mathcal{U}} \sum_{m \in \mathcal{E}} ((\boldsymbol{\breve{p}}_u^m)^{\top} \cdot \boldsymbol{\tilde{p}}_u^m)^2,
\end{equation}
\begin{equation}
    \label{eq:bhv_emb}
    \boldsymbol{\breve{p}}_u^m = \boldsymbol{p}_u^m - \boldsymbol{\tilde{p}}_u^m,
\end{equation}
where $|\cdot|$ means the length of the set. $\boldsymbol{p}_u^m$ represents the user embeddings (i.e., the input of the encoder) learned in the $m$-th environment, $\boldsymbol{\tilde{p}}_u^m$ and $\boldsymbol{\breve{p}}_u^m$ denote the user invariant preferences and behavior-specific preferences for the $m$-th environment, respectively.

\textbf{\textit{Contrastive Loss}}.
Given that there should be a similarity between the invariant preferences of the same user across different environments (see Fig.~\ref{fig:toys}(d)), we introduce contrastive learning to enforce constraints on the learned invariant preferences. Specifically, we treat the pair of two invariant preferences as a positive sample pair, and an invariant preference and behavior-specific preference pair as a negative sample pair. The loss function adopts InfoNCE~\cite{InfoNCE} loss:
\begin{equation}
    \label{eq:nce}
    \mathcal{L}_{con} = - \sum_{u \in \mathcal{U}} \sum_{m, n \in \mathcal{E}} log \frac{exp((\boldsymbol{\tilde{p}}_u^m \cdot \boldsymbol{\tilde{p}}_u^n) /\tau)}{exp((\boldsymbol{\tilde{p}}_u^m \cdot \boldsymbol{\tilde{p}}_u^n) /\tau) + \sum_{j \in \mathcal{E}} exp((\boldsymbol{\tilde{p}}_u^m \cdot \boldsymbol{\breve{p}}_u^j) /\tau) },
\end{equation}
where $\tau$ is a hyperparameter used to adjust the similarity scale, $\boldsymbol{\breve{p}}_u^j$ represents an element in the set of user behavior-specific preferences, which is calculated by Eq.~\ref{eq:bhv_emb}.

\textbf{\textit{Invariant Preference Generation Loss}}.
The three loss functions mentioned above are jointly optimized to generate accurate user invariant preferences. The corresponding invariant preference generation loss is expressed as:
\begin{equation}
    \label{eq:inv}
    \mathcal{L}_{inv} = \mathcal{L}_{irm} + \mathcal{L}_{ort} +  \mathcal{L}_{con} + \mathcal{L}_{kl},
\end{equation}
where $\mathcal{L}_{kl}=\boldsymbol{D}_{kl}(\mathcal{N}(\boldsymbol{\mu}, \boldsymbol{\sigma}^2) || \mathcal{N}(0, 1))$ is used to compute the KL divergence of $\mathcal{N}(\boldsymbol{\mu}, \boldsymbol{\sigma}^2)$ and $\mathcal{N}(0, 1)$, which is a common regularization term in VAEs~\cite{VAE} to prevent the decoder from sampling noise as 0. 

\subsection{Recommendation}
To better provide recommendations for the target behavior, we aggregate the learned invariant preferences from all environments as the final user invariant preference to achieve complementary information. The final user invariant preference is:
\begin{equation}
  \label{eq:agg_p}
   \boldsymbol{\tilde{p}}_u^{agg} = \sum_{m \in \mathcal{E}} \boldsymbol{\tilde{p}}_u^m,
\end{equation}
where $\boldsymbol{\tilde{p}}_u^{agg}$ denotes the final user invariant preference, and $\mathcal{E}$ represents the environment set.

It is important to note that behavior-specific preferences represent a user's preferences for different external influences in a specific behavior. This means the target behavior-specific preferences are crucial for accurately predicting the target behavior. Thus, we consider both the user's invariant preferences and the target behavior-specific preference to provide recommendations for the user. Following previous CF-based approaches~\cite{MBGCN, CRGCN, BCIPM}, we calculate the user-item correlation for recommendations:
\begin{equation}
    \label{eq:pred_score}
    y_{ui} = (\boldsymbol{\tilde{p}}_u^{agg})^{\top} \cdot \boldsymbol{\hat{q}}_i + (\boldsymbol{\breve{p}}_u^{tar})^{\top} \cdot \boldsymbol{\hat{q}}_i,
\end{equation}
where $\boldsymbol{\hat{q}}_i$ represents the embedding of item $i$, learned by Eq.~\ref{eq:sum_q}, and $\boldsymbol{\breve{p}}_u^{tar}$ denotes the target behavior-specific preference of user $u$, obtained through Eq.~\ref{eq:bhv_emb}.

We optimize the recommendation task using the bayesian personalized ranking (BPR) loss function:
\begin{equation}
  \label{eq:bpr}
  \mathcal{L}_{rec} = -\sum_{(u,i,j) \in \mathcal{O}} ln \sigma(y_{ui}-y_{uj}),
\end{equation}
where $\sigma (\cdot)$ denotes the sigmoid function, $\mathcal{O}=\{(u,i,j) \in \mathcal{R}|r_{ui}=1, r_{uj}=0\}$ represents the observed ($r_{ui} = 1$) and unobserved ($r_{uj} = 0$) interactions in the target behavior matrix  $\mathcal{R}$.

Finally, we jointly optimize the invariant preference generation loss and the recommendation loss. The total loss function is formulated as follows:
\begin{equation}
  \label{eq:total}
  \mathcal{L} = \mathcal{L}_{rec} + \mathcal{L}_{inv},
\end{equation}
where we omit the regularization term. To facilitate the discussion of hyperparameters, we rewrite it as: 
\begin{equation}
  \label{eq:total_final}
  \mathcal{L} = \mathcal{L}_{rec} + \lambda \cdot \mathcal{L}_{irm} + \alpha \cdot \mathcal{L}_{ort} + \beta \cdot \mathcal{L}_{con} + \gamma \cdot \mathcal{L}_{kl},
\end{equation}
where $\lambda, \alpha, \beta, \gamma$ are hyperparameters.
\section{Experiment}
In this section, we present a series of experiments designed to validate the effectiveness of our model and analyze the rationale behind its design.

\subsection{Experiment Settings}
\subsubsection{\textbf{Dataset}}
This study leverages four real-world datasets:
\begin{itemize}[leftmargin=*]
\item \textbf{Yelp}\footnote{https://www.kaggle.com/yelp-dataset/yelp-dataset}. This dataset includes tips and ratings. Previous studies~\cite{GNMR, KHGT} categorized ratings into three behaviors based on score. In Yelp, \textit{like} is the target behavior, and \textit{tips}, \textit{dislike}, and \textit{neutral} are auxiliary behaviors.

\item \textbf{ML10M}\footnote{https://grouplens.org/datasets/movielens/10m/}. ML10M comes from the GroupLens project for movie recommendation. This is a ratings dataset. Similar to Yelp, the dataset is categorized into three types of behaviors: \textit{like}, \textit{dislike}, and \textit{neutral}, where \textit{like} represents the target behavior, and \textit{dislike} and \textit{neutral} represent auxiliary behaviors.

\item \textbf{Taobao}\footnote{https://tianchi.aliyun.com/dataset/649}. This dataset is sourced from an e-commerce platform under the Alibaba Group. It comprises three types of behaviors, where \textit{purchase} represents the target behavior, and \textit{click} and \textit{cart} represent auxiliary behaviors.

\item \textbf{Tmall}\footnote{https://tianchi.aliyun.com/dataset/140281}. This dataset originates from another e-commerce platform operated under the Alibaba Group. It includes four types of behaviors, with \textit{purchase} denoting the target behavior, while \textit{click}, \textit{collect}, and \textit{cart} are auxiliary behaviors.

\end{itemize}

Following established practices in prior research~\cite{DGCF, CRGCN}, we retain only the initial occurrence of each interaction to eliminate redundant data. For further details about the dataset, please refer to Tab.~\ref{tab:dataset}.

\begin{table}[htb]
  \caption{Statistics of four datasets ("-" means without this behavior).}
  \label{tab:dataset}
    \begin{tabular}{cccccccc}
      \toprule
      \textbf{Dataset} & \textbf{\# User} & \textbf{\# Item} & \textbf{\# click}  & \textbf{\# collect} & \textbf{\# cart} & \textbf{\# purchase} \\
      \midrule
      \textbf{Taobao} & 48,749 & 39,493 & 1,548,162 &       - & 193,747 & 211,022\\
      \textbf{Tmall}  & 41,738 & 11,953 & 1,813,498 & 221,514 &  1,996  & 255,586 \\
      \bottomrule
      \toprule
      \textbf{Dataset} & \textbf{\# User} & \textbf{\# Item} & \textbf{\# tips}  & \textbf{\# dislike} & \textbf{\# neutral} & \textbf{\# like} \\
      \midrule
      \textbf{Yelp}   & 19,800 & 22,734 & 285,673   & 198,106 & 238,880 & 657,926 \\
      \textbf{ML10M}  & 67,788 &  8,704 &       -   & 1,370,897 & 3,580,155 & 4,970,984 \\
      \bottomrule
    \end{tabular}
\end{table}

\subsubsection{\textbf{Evaluation Protocols}}
We employ the widely used leave-one-out approach~\cite{MBGCN, NMTR} for model evaluation. Specifically, we set aside the last interaction of each user for testing, while using the remaining interactions as training data. In the test phase, we predict all use-item interactions that do not occur in the training set, and then evaluate top@k for performance. We also utilize two evaluation metrics, \emph{Hit Ratio (HR@K)}~\cite{HR} and \emph{Normalized Discounted Cumulative Gain (NDCG@K)}~\cite{NDCG}, to assess the model's effectiveness. \emph{HR@K} measures whether the recommendation list contains items that the user actually interacts with, while \emph{NDCG@K} considers the relationship between the ranking of recommended items and their relevance to the user.

\subsubsection{\textbf{Parameter Settings}}
We set the embedding size to 64 and the batch size to 1,024. The number of GCN layers is uniformly set to 2. Our model utilizes the Adam~\cite{ADAM} optimizer for gradient descent and the learning rate is set to $1e^{-3}$. For the hyperparameters $\lambda$, $\alpha$, $\beta$, $\gamma$, and $\tau$, we perform a grid search to find the optimal parameters from the sets $\{ 0.01, 0.1, 0.5, 1.0 \}$, $\{ 1e^{-4}, 1e^{-3}, 1e^{-2}, 0.1, 1.0 \}$, $\{ 1e^{-3}, 1e^{-2}, 0.1, 0.5, 1.0 \}$, $\{ 0.1, 0.3, 0.5, 0.7, 1.0 \}$, and $\{ 0.1, 0.3, 0.5, 0.7, 0.9 \}$, respectively.

\subsubsection{\textbf{Baselines}}
To effectively evaluate the performance of our model, we compare it with two representative single-behavior and seven representative multi-behavior methods.

\noindent \textbf{Single-behavior models:}

\begin{itemize}[leftmargin=*]
  \item \textbf{MF-BPR}~\cite{BPR}. This method utilizes matrix factorization for recommendation and optimizes using the bayesian personalized ranking (BPR) loss.
  \item \textbf{LightGCN}~\cite{LightGCN}. It simplifies graph convolutional network (GCN) operations, enhancing performance while reducing model complexity.
\end{itemize}

\noindent \textbf{Multi-behavior models:}

\begin{itemize}[leftmargin=*]
  \item \textbf{NMTR}~\cite{NMTR}. NMTR establishes dependencies between behaviors based on their temporal order of occurrence.
  \item \textbf{MBGCN}~\cite{MBGCN}. The model learns user preferences within each behavior graph and aggregates them considering the importance of each behavior.
  \item \textbf{S-MBRec}~\cite{SMBREC}. The model incorporates a star-structured contrastive learning module to explore commonalities between different behaviors.
  \item \textbf{CRGCN}~\cite{CRGCN}. CRGCN models dependencies between behaviors by sequentially propagating embeddings across different behaviors, refining user preferences iteratively.
  \item \textbf{MB-GCGN}~\cite{MB-CGCN}. This model extends CRGCN by optimizing the model structure and aggregating user preferences from all behaviors for recommendation.
  \item \textbf{PKEF}~\cite{PKEF}. This model integrates a parallel knowledge fusion module and a projection disentangling multi-expert network to address data distribution imbalance.
  \item \textbf{BCIPM}~\cite{BCIPM}. BCIPM designs a network and incorporates auxiliary behavior interaction data to facilitate the training of this network. This network is then used to extract the item-aware information from the target behavior interactions for recommendations, effectively preventing the noise introduced by auxiliary behaviors.
\end{itemize}

\subsection{Overall Performance}

Tab.~\ref{tab:overall} presents a comparison of our model with baseline methods on the HR and NDCG metrics for the top@10 results across four datasets. Analyzing the experimental results, we can draw the following observations:

\begin{table*}[htb]
  \caption{Overall performance across four datasets ("\textit{Impr.}" represents the relative performance improvement compared to the suboptimal metric, the bold font indicates the optimal metric, the underline represents the suboptimal metric, and "*" denotes statistically significant improvements, as determined by two-tailed t-test with $p < 0.05$ ).}
  \label{tab:overall}
  \resizebox{\textwidth}{!}{
	\begin{tabular}{crccccccccccr}
	\toprule
	\multirow{2}{*}[-2ex]{\textbf{Dataset}} & \multirow{2}{*}[-2ex]{\makecell[c]{\textbf{Metric} \\ (\textit{top@10})}} & \multicolumn{2}{c}{\textbf{Single-behavior}}       & \multicolumn{8}{c}{\textbf{Multi-behavior}}    & \multirow{2}{*}[-2ex]{\textit{\textbf{Impr.}}} \\ \cmidrule(lr){3-4} \cmidrule(lr){5-12}
	                                             &                     & \makecell[c]{\textbf{MF-BPR} \\ \textit{UAI'09}} & \makecell[c]{\textbf{LightGCN} \\ \textit{SIGIR'20}}  & \makecell[c]{\textbf{NMTR} \\ \textit{TKDE'21}} & \makecell[c]{\textbf{MBGCN} \\ \textit{SIGIR'20}} & \makecell[c]{\textbf{S-MBRec} \\ \textit{IJCAI'22}} & \makecell[c]{\textbf{CRGCN} \\ \textit{TOIS'23}} & \makecell[c]{\textbf{MB-CGCN} \\ \textit{WWW'23}} & \makecell[c]{\textbf{PKEF} \\ \textit{CIKM'23}} & \makecell[c]{\textbf{BCIPM} \\ \textit{SIGIR'24}}  & \textbf{UIPL}  & \\ \midrule  
	\multirow{2}{*}{\textbf{Tmall}}   & \textbf{HR}   & 0.0230          & 0.0393           & 0.0517         & 0.0549        & 0.0694         & 0.0840        & 0.1073     & 0.1118 & \underline{0.1414}  & $\boldsymbol{0.1434}^*$ & 1.41\%  \\ 
					  & \textbf{NDCG} & 0.0124          & 0.0209             & 0.0250         & 0.0285        & 0.0362         & 0.0442        & 0.0416     & 0.0630 & \underline{0.0741}  & $\boldsymbol{0.0790}^*$ & 6.61\%  \\ \midrule

	\multirow{2}{*}{\textbf{Taobao}}  & \textbf{HR}   & 0.0178          & 0.0254           & 0.0409         & 0.0434        & 0.0571         & 0.1152        & 0.0989     & 0.1097 & \underline{0.1292}  & $\boldsymbol{0.1328}^*$ & 2.79\%  \\ 
					  & \textbf{NDCG} & 0.0101          & 0.0138           & 0.0212         & 0.0259        & 0.0331         & 0.0629        & 0.0470     & 0.0627 & \underline{0.0716}  & $\boldsymbol{0.0782}^*$ & 9.22\%  \\ \midrule

	\multirow{2}{*}{\textbf{Yelp}}    & \textbf{HR}   & 0.0327          & 0.0400           & 0.0324         & 0.0356        & 0.0353         & 0.0367        & 0.0355     & 0.0423 & \underline{0.0502}  & $\boldsymbol{0.0522}^*$ &  3.98\%  \\ 
					  & \textbf{NDCG} & 0.0159          & 0.0202           & 0.0160         & 0.0183        & 0.0173         & 0.0178        & 0.0164     & 0.0210 & \underline{0.0244}  & $\boldsymbol{0.0254}^*$ &  4.10\%  \\ \midrule

	\multirow{2}{*}{\textbf{Ml10M}}   & \textbf{HR}   & 0.0585          & 0.0662           & 0.0451         & 0.0469        & 0.0340         & 0.0502        & 0.0627     & 0.0591 & \underline{0.0810}  & $\boldsymbol{0.0835}^*$ & 3.09\%  \\ 
					  & \textbf{NDCG} & 0.0274          & 0.0319           & 0.0202         & 0.0228        & 0.0164         & 0.0239        & 0.0302     & 0.0264  & \underline{0.0392}  & $\boldsymbol{0.0401}^*$ & 2.30\%  \\ 
	\bottomrule
	\end{tabular}
  }

\end{table*}

(1) Across the four datasets, our model consistently outperforms the baseline methods in terms of HR@10 and NDCG@10 metrics, demonstrating the effectiveness of our proposed method. This improvement can be attributed to our approach of modeling user invariant preferences, where we decompose user preferences into invariant and behavior-specific components. By introducing the invariant risk minimization paradigm and constructing multiple environments for learning user invariant preferences, UIPL effectively mitigates the introduction of noise. 
It is worth noting that, we observe a significant difference in the average relative improvement of the HR@10 metric on the Tmall dataset compared to the other three datasets. We attribute this result to the manner we used to augment the environment set. Specifically, we apply a union operation to combine data from different behaviors to expand the environment set (see Section~\ref{erl}). However, in the Tmall dataset, the interaction data for the “cart” behavior is very sparse (see Tab.~\ref{tab:dataset}), causing some environments in the expanded set to be ineffective for a large number of users, which in turn affects the HR@10 metric. Additionally, the performance improvement of UIPL on the Yelp and ML10M datasets is not particularly pronounced, especially regarding the NDCG@10 metric. This can be attributed to the nature of the behavioral relationships. The correlation between behaviors in the Yelp and ML10M datasets is relatively low, leading to less informative learned invariant preferences. To address this, we can explore two potential avenues for enhancing the expressiveness of invariant preferences: developing more effective environment construction strategies and improving the generative capacity of the VAE network.

(2) On the overall view, multi-behavior methods outperform single-behavior methods because of their ability to model diverse behavioral data and capture rich user preferences in auxiliary behaviors, effectively mitigating the data sparsity problem. Among the two single-behavior methods, LightGCN outperforms MF-BPR, indicating the superior effectiveness of modeling all interactions compared to modeling individual interactions. LightGCN performs information propagation and aggregation within the graph, significantly enriching the node representations, a capability that MF-BPR method lacks. Interestingly, most multi-behavior methods do not exhibit substantial improvements over single-behavior methods in the Yelp and ML10M datasets. We attribute this to low inter-behavior correlations in these datasets. Since most methods heavily rely on inter-behavior correlations, this results in poorer performance when such correlations are weak.

(3) Among the multi-behavior baseline methods, NMTR solely focuses on modeling the cascading influence between multiple behaviors, resulting in low performance. MBGCN and S-MBRec individually model each behavior and then perform weighted aggregation. S-MBRec utilizes a more intricate weighting scheme than MBGCN, leading to superior performance in datasets with higher inter-behavior correlations, such as Tmall and Taobao. However, these methods do not exhibit significant improvements on the Yelp and MB10M datasets, highlighting the limitation of the weighted aggregation method. In contrast, CRGCN, MB-CGCN, and PKEF adopt a sequential modeling approach, facilitating the aggregation of information between behaviors through embedding propagation. They outperform MBGCN and S-MBRec, indicating that sequential modeling to explore relationships between behaviors is more effective than weighted aggregation based on inter-behavior correlations. BCIPM recognizes the presence of noise in auxiliary behaviors and therefore designs a specialized network to extract information from interactions. Data from auxiliary behaviors is exclusively used to train this network, effectively avoiding the noise issues that may arise from incorporating auxiliary behavior information. The performance of this method outperforms that of other baseline models, highlighting the importance of addressing noise-related challenges.

Next, we will conduct a series of ablation studies to analyze and validate the effectiveness and rationale of each component in our model design.

\subsection{Ablation Study}
Our model consists of three components: embedding learning, invariant preference generation, and recommendation. To validate the effectiveness of each module, we conducted an ablation analysis from the following perspectives.

\subsubsection{\textbf{Embedding Pretraining Analysis}}
\begin{figure}[ht]
    \centering
    \includegraphics[width=0.75\linewidth]{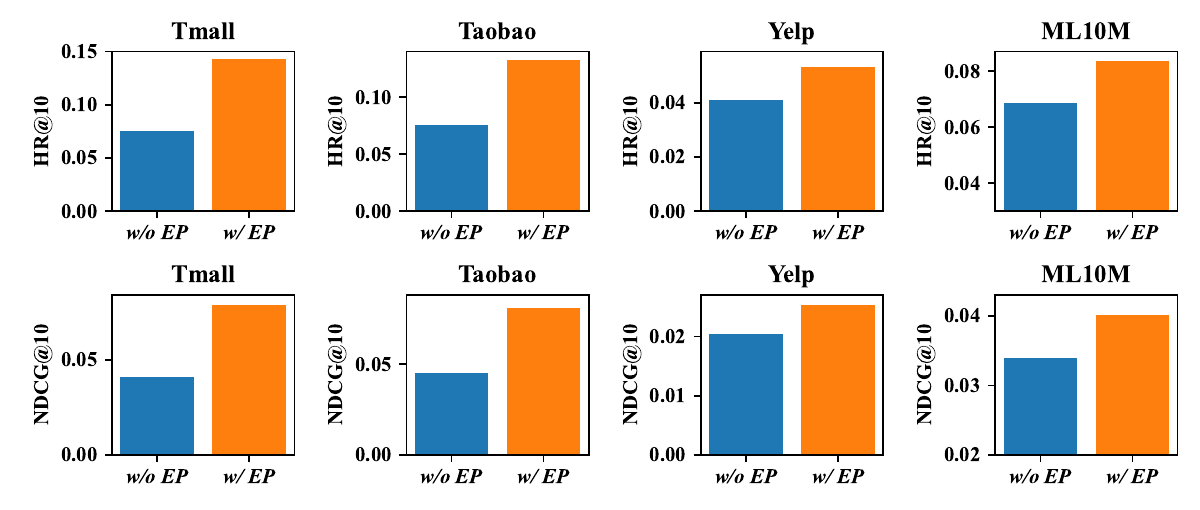}
    \caption{Comparison between with and without embedding pretraining ("\textit{\textbf{w/ EP}}" represents with embedding pretraining and "\textit{\textbf{w/o EP}}" means without embedding pretraining).}
    \label{fig:pretrain}
    \Description{}
\end{figure}

Before conducting environmental representation learning, we pretrain the embeddings of users and items, which are initialized randomly. To assess the effectiveness of this design, we perform an ablation study on the embedding pretraining module. The experimental settings are as follows: (1) \textit{\textbf{w/o EP}}: excludes the embedding pretraining module, where randomly initialized embeddings are directly used as inputs to the environmental representation learning module. (2) \textit{\textbf{w/ EP}}: includes the embedding pretraining module. The experimental results are shown in Fig.~\ref{fig:pretrain}.

The results indicate that, across the four datasets, the method that includes embedding pretraining significantly outperforms the method without pretraining, highlighting the importance of this step. Without pretraining, the model directly shares randomly initialized embeddings within downstream modules, which can lead to conflicts in the information and hinder effective feature extraction. 

In contrast, by incorporating pretraining, the model first constructs a global interaction graph using data from all behaviors. It then applies GCN for information propagation and aggregation on this graph, producing coarse-grained user representations. These representations are subsequently used as initial embeddings for downstream modules, where further operations in the environmental representation learning module can be viewed as performing graph convolution on subgraphs. This process refines the pretrained embeddings locally, helping to prevent information conflicts. As a result, it mitigates the challenges downstream modules face in extracting meaningful features.

\subsubsection{\textbf{Environment Augmentation Analysis}} \label{env_analy}
\begin{figure}[ht]
    \centering
    \includegraphics[width=0.75\linewidth]{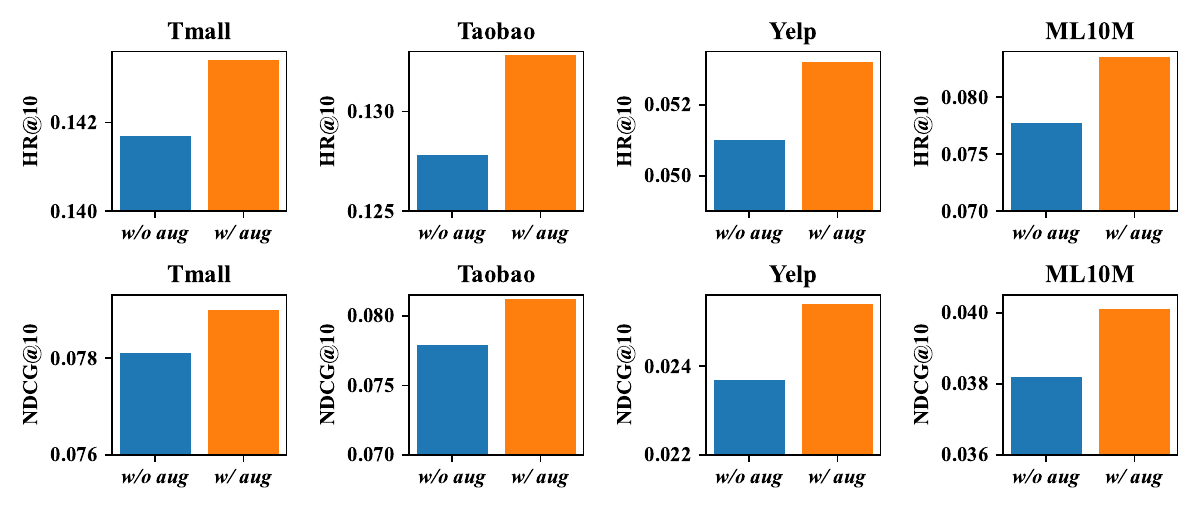}
    \caption{Environment augmentation analysis ("\textit{\textbf{w/ aug}}" indicates an augmented environment set, while "\textit{\textbf{w/o aug}}" denotes no augmentation of the environment set).}
    \label{fig:combine}
    \Description{}
\end{figure}

IRM aims to learn features that remain stable across different environments by simulating potential distribution shifts. To strengthen the constraint of IRM, we augment the environment set by generating all combinations of user behaviors. We conduct the following experiment to evaluate the effectiveness of this approach: (1) \textit{\textbf{w/o aug}}: without expanding the environment set, using only individual behavior sets as the environment. (2) \textit{\textbf{w/ aug}}: generating all combinations of user behaviors to form the environment set. The experimental results are shown in Fig.~\ref{fig:combine}.

From the result, we observe a significant improvement in model performance across the four datasets following the enhancement of the environment. This result validates the effectiveness of our environment augmentation strategy. The expansion of the environment set provides stronger support for the IRM constraint. By combining different behaviors, we effectively simulate a range of real-world scenarios, introducing diverse environments that help the model learn more general and invariant features.

The lack of sufficient environmental diversity may hinder the model's ability to generalize effectively. Therefore, it is crucial to construct an adequate and diverse set of environments to facilitate the learning of user invariant preferences.

\subsubsection{\textbf{Loss Function Analysis}}

Our method employs a joint optimization approach incorporating both the recommendation loss and the invariant preference generation loss. The invariant preference generation loss includes several auxiliary loss functions designed to facilitate the extraction of invariant preferences. To assess the contribution of each loss function, we conduct an ablation study with the following configurations: (1) \textbf{\textit{$\mathcal{L}_{rec}$ only}}: remove the invariant preference generation loss, retaining only the recommendation loss. (2) \textbf{\textit{w/o $\mathcal{L}_{kl}$}}: remove the KL divergence loss. (3) \textbf{\textit{w/o $\mathcal{L}_{irm}$}}: remove the IRM loss. (4) \textbf{\textit{w/o $\mathcal{L}_{ort}$}}: remove the orthogonal loss. (5) \textbf{\textit{w/o $\mathcal{L}_{con}$}}: remove the contrastive loss. The experimental results are presented in Tab.~\ref{tab:amb}.

\begin{table}[ht]
 \caption{Statistics of loss function ablation experiment results ("\textbf{\textit{$\mathcal{L}_{rec}$ only}}" represents removing the invariant preference generation loss and retaining only the recommendation loss. "\textbf{\textit{w/o $\mathcal{L}_{kl}$}}", "\textbf{\textit{w/o $\mathcal{L}_{irm}$}}", "\textbf{\textit{w/o $\mathcal{L}_{ort}$}}", "\textbf{\textit{w/o $\mathcal{L}_{con}$}}" respectively represent individually removing KL divergence loss, IRM loss, orthogonal loss, and contrastive loss).}
\label{tab:amb}
\resizebox{0.85\textwidth}{!}{
\begin{tabular}{rcccccccc}
\toprule
\multirow{2}{*}{\textbf{Method}} & \multicolumn{2}{c}{\textbf{Tmall}} & \multicolumn{2}{c}{\textbf{Taobao}} & \multicolumn{2}{c}{\textbf{Yelp}} & \multicolumn{2}{c}{\textbf{ML10M}} \\ \cmidrule(lr){2-3} \cmidrule(lr){4-5} \cmidrule(lr){6-7} \cmidrule(lr){8-9}

                                 & \multicolumn{1}{c}{\textbf{HR@10}} & \multicolumn{1}{c}{\textbf{NDCG@10}} & \multicolumn{1}{c}{\textbf{HR@10}} & \multicolumn{1}{c}{\textbf{NDCG@10}} & \multicolumn{1}{c}{\textbf{HR@10}} & \multicolumn{1}{c}{\textbf{NDCG@10}} & \multicolumn{1}{c}{\textbf{HR@10}} & \multicolumn{1}{c}{\textbf{NDCG@10}} \\ \midrule  
\textbf{\textit{$\mathcal{L}_{rec}$ only}}  & 0.1272 & 0.0649 & 0.1203 & 0.0695 & 0.0416 & 0.0209 & 0.0534 & 0.0258 \\ \midrule
\textbf{\textit{w/o $\mathcal{L}_{kl}$}}    & 0.1309 & 0.0685 & 0.1305 & 0.0797 & 0.0495 & 0.0240 & 0.0584 & 0.0284 \\ \midrule
\textbf{\textit{w/o $\mathcal{L}_{irm}$}}   & 0.1311 & 0.0774 & 0.1302 & 0.0797 & 0.0508 & 0.0239 & 0.0653 & 0.0319 \\ \midrule  
\textbf{\textit{w/o $\mathcal{L}_{ort}$}}   & 0.1403 & 0.0770 & 0.1319 & 0.0802 & 0.0511 & 0.0242 & 0.0686 & 0.0340 \\ \midrule  
\textbf{\textit{w/o $\mathcal{L}_{con}$}}   & 0.1406 & 0.0773 & 0.1282 & 0.0786 & 0.0497 & 0.0238 & 0.0716 & 0.0351 \\ \midrule  
\textbf{UIPL}      & \textbf{0.1434} & \textbf{0.0790} & \textbf{0.1328} & \textbf{0.0812} & \textbf{0.0532} & \textbf{0.0254} & \textbf{0.0835} & \textbf{0.0401}      \\
\bottomrule               
\end{tabular}
}
\end{table}

The experimental results indicate that removing any of the loss functions leads to a significant decrease in model performance across all four datasets. This finding underscores the effectiveness of each loss function in our method. Specifically, removing the invariant preference generation loss compromises the model's ability to constrain user invariant preferences. As a result, behavior-specific preferences from auxiliary behaviors are introduced when recommending for target behaviors, which hinders the recommendation process. By constraining the invariant preferences, the model can learn user preferences that remain consistent across different environments and integrate them during the recommendation process, enhancing performance through complementary information. 

The KL divergence loss quantifies the difference between the latent space distribution output by the encoder and the prior distribution, encouraging the model to learn structured latent representations. Without this loss, the learned latent representations lose their structure, making it difficult to accurately capture the data distribution, ultimately degrading the quality of the generated invariant preferences. We observe that the ML10M dataset is more significantly affected by the absence of the KL divergence loss than the other three datasets. This may be due to the higher complexity of the ML10M dataset, which is more sensitive to structured latent representations.

The IRM loss, orthogonal loss, and contrastive loss collectively constrain the generation of user invariant preferences. Among them, the IRM loss has a greater impact on model performance than the other two. This is because the IRM loss utilizes prior information from multiple environments to constrain predictions, whereas the orthogonal and contrastive losses provide more localized constraints on the user’s embeddings, leading to a smaller impact on the overall performance.

In this section, we validate the effectiveness of each loss function through ablation experiments. However, a detailed analysis of the contribution of each individual loss function is lacking. Our method employs multi-task learning for joint optimization, and to further understand the role of each loss function, we will quantitatively analyze their contributions from a hyperparameter perspective in the next section.

\subsubsection{\textbf{Invariant Preferences Analysis}}

\begin{figure}[ht]
    \centering
    \includegraphics[width=0.9\linewidth]{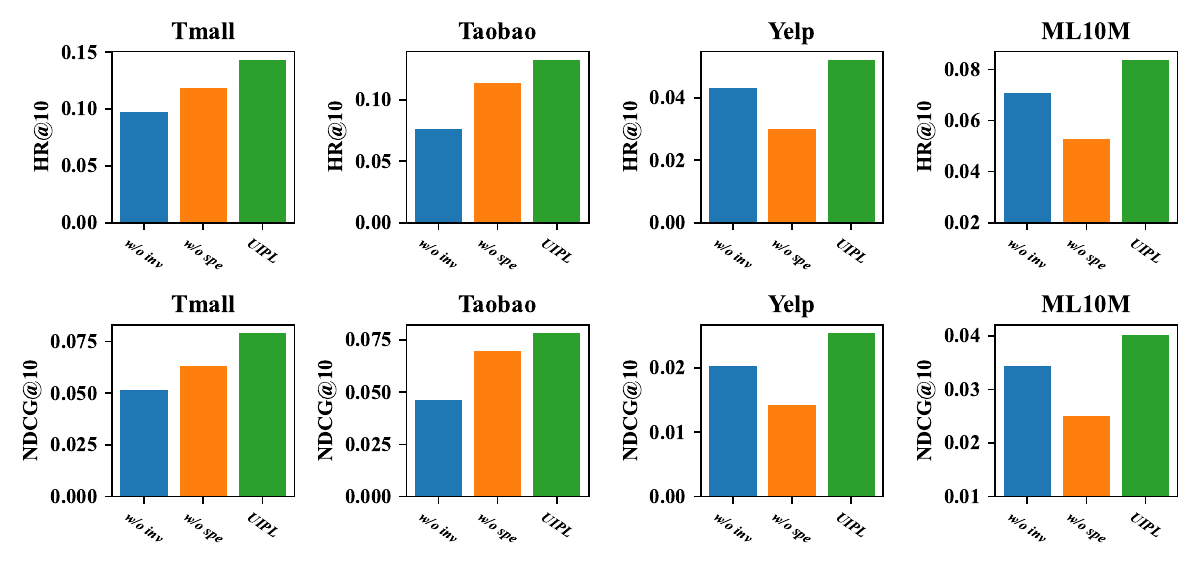}
    \caption{Analysis of invariant preferences ("\textit{\textbf{w/o inv}}" represents the removal of invariant preferences, "\textit{\textbf{w/o spe}}" represents the removal of target-behavior-specific preferences).}
    \label{fig:inv}
    \Description{}
\end{figure}

Invariant preferences learned from multi-behavior interactions are essential for enhancing the model’s performance. To further evaluate the effectiveness of invariant preferences, we conduct the following experiments: (1) \textit{\textbf{w/o inv}}: remove invariant preferences and only retain behavior-specific preferences learned from the target behavior for recommendations. (2) \textit{\textbf{w/o spe}}: remove behavior-specific preferences learned from the target behavior and rely solely on invariant preferences for recommendations. We compare the results of the two experiments mentioned above with the method UIPL, which includes both invariant preferences and target-behavior-specific preferences.  The experimental results are presented in Fig.~\ref{fig:inv}.

From the experimental results, it can be observed that the UIPL method, which includes both invariant preferences and target-behavior-specific preferences, achieves the best performance on all four datasets. This further validates the effectiveness of the UIPL design, demonstrating that both types of preferences significantly impact recommendation performance. When comparing the results of removing invariant preferences and target-behavior-specific preferences, we observe the following trends: on the Tmall and Taobao datasets, the model performs better without target-behavior-specific preferences (\textit{\textbf{w/o spe}}) than without invariant preferences (\textit{\textbf{w/o inv}}). In contrast, on the Yelp and ML10M datasets, the reverse is true. This variation can be attributed to the differing correlations between behaviors in the datasets. Specifically, the correlations between behaviors in the Tmall and Taobao datasets are stronger than those in the Yelp and ML10M datasets, allowing the model to learn more invariant preferences on the former. Conversely, target-behavior-specific preferences play a more substantial role in recommendation performance on the Yelp and ML10M datasets. These findings suggest that adopting a strategy that effectively combines both types of preferences could further enhance the model’s performance, which will be a focus of our future research.

\subsubsection{\textbf{Hyperparameter Analysis}}
In UIPL, we define hyperparameters to control the contribution of each loss function. To further analyze the model’s effectiveness, we employ a controlled variable approach to analyze the impact of each hyperparameter. The specific experimental setup and the corresponding results are provided below.

\begin{figure}[ht]
    \centering
    \includegraphics[width=0.9\linewidth]{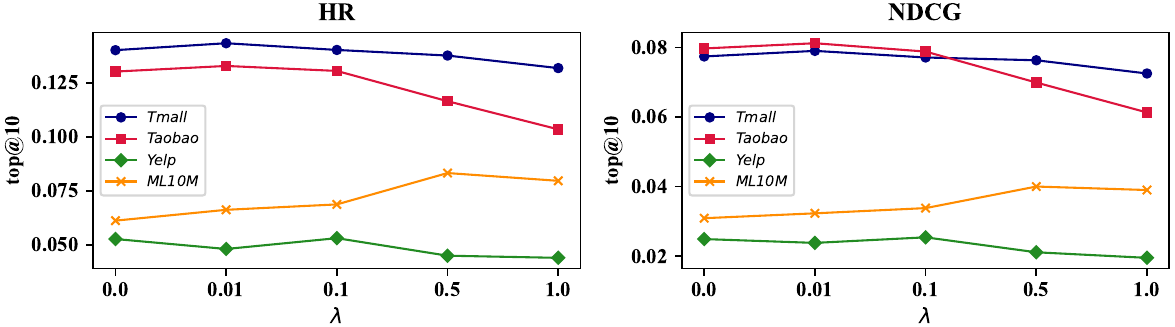}
    \caption{Analysis of the hyperparameter $\lambda$ (x-axis represents the values of $\lambda$, and y-axis represents the model's performance).}
    \label{fig:lambda}
    \Description{}
\end{figure}

\textit{\textbf{Hyperparameter $\boldsymbol{\lambda}$}}. 
$\boldsymbol{\lambda}$ is used to control the contribution of the IRM loss. We record the performance of the model with different values of $\boldsymbol{\lambda}$. The experimental results are shown in Fig.~\ref{fig:lambda}.

From the experimental results, it is evident that as the value of $\lambda$ increases, the overall performance of the model across the four datasets initially improves and then declines. The optimal values of $\lambda$ vary across the datasets, reflecting differences in data distribution and complexity.

In the Tmall dataset, the performance curve exhibits gradual changes with variations in $\lambda$, suggesting that the model is relatively less sensitive to the IRM loss in this dataset. In the Yelp dataset, we observe a slight decline in model performance as the $\lambda$ increases from 0.0 to 0.01. This could be attributed to the model reaching a local optimum. For the ML10M dataset, the model's performance deteriorates only after the $\lambda$ exceeds 0.5, indicating the higher importance of the IRM loss in the ML10M dataset.

\begin{figure}[ht]
    \centering
    \includegraphics[width=0.9\linewidth]{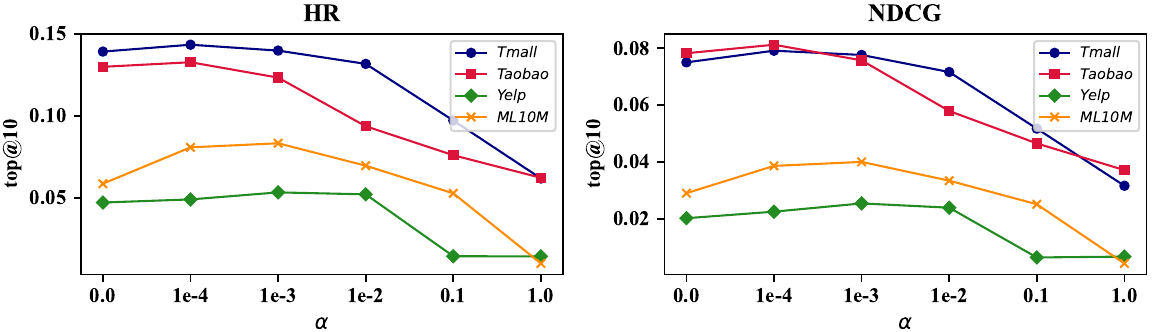}
    \caption{Analysis of the hyperparameter $\alpha$ (x-axis represents the values of $\alpha$, and y-axis represents the model's performance).}
    \label{fig:alpha}
    \Description{}
\end{figure}

\textit{\textbf{Hyperparameter $\boldsymbol{\alpha}$}}.
The orthogonal loss is designed to preserve the orthogonality between user invariant preferences and behavior-specific preferences, thereby minimizing their correlation. $\boldsymbol{\alpha}$ is a hyperparameter associated with the orthogonal loss. To assess its influence on model performance, we vary its value between 0 and 1 to explore the effects of different orders of magnitude. The experimental results are presented in Fig.~\ref{fig:alpha}.

The experimental results indicate that the model's performance improves gradually with increasing values of $\alpha$ across all four datasets, but sharply declines once $\alpha$ exceeds a certain threshold. When $\alpha$ surpasses 0.1, performance deteriorates noticeably on all datasets, particularly on the Yelp and ML10M datasets. This phenomenon can be attributed to the orthogonal loss having a significant magnitude. With smaller values of $\alpha$, the orthogonal loss helps optimize the model. However, as $\alpha$ increases, the model tends to minimize the orthogonal loss during training, which detracts from the optimization of other tasks and leads to a significant decline in recommendation performance.

\begin{figure}[ht]
    \centering
    \includegraphics[width=0.9\linewidth]{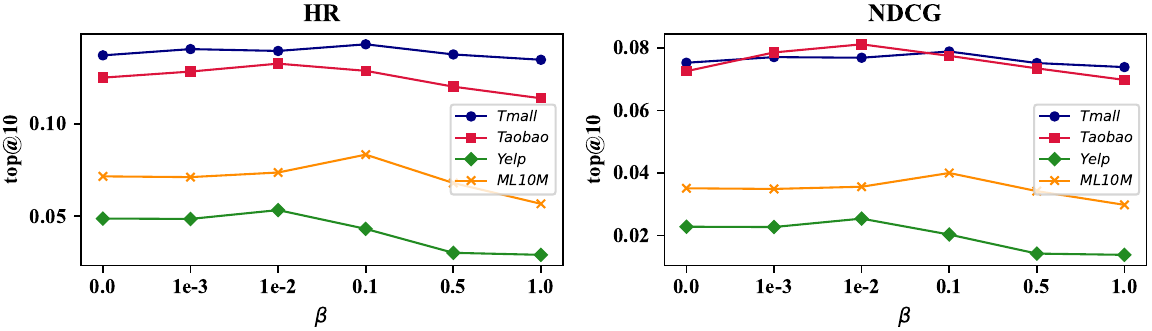}
    \caption{Analysis of the hyperparameter $\beta$ (x-axis represents the values of $\beta$, and y-axis represents the model's performance).}
    \label{fig:beta}
    \Description{}
\end{figure}

\textit{\textbf{Hyperparameter $\boldsymbol{\beta}$}}. 
$\boldsymbol{\beta}$ is a hyperparameter assigned to the contrastive loss, which aims to reduce the correlation between the invariant preferences generated by the decoder across different environments, thereby facilitating the model's optimization for invariant preference generation. To further investigate the contribution of this constraint, we compare the influence of varying $\boldsymbol{\beta}$ values on model performance. The experimental results are presented in Fig.~\ref{fig:beta}.

Overall, the model's performance exhibits a trend of initially rising and then declining with increasing $\boldsymbol{\beta}$ values. On the Tmall and Taobao datasets, the performance curves show relatively smooth variations with changes in $\boldsymbol{\beta}$, while on the Yelp and ML0M datasets, the changes are more pronounced. This phenomenon can be attributed to differences in behavioral correlations within the datasets. The Tmall and Taobao datasets exhibit higher correlations between behaviors, resulting in a smaller influence on data distribution after constructing combinations of different contexts and a higher correlation between the generated invariant preferences. In contrast, the performance on Yelp and ML0M is the opposite, thus these datasets are more sensitive to the contrastive loss.

\begin{figure}[ht]
    \centering
    \includegraphics[width=0.9\linewidth]{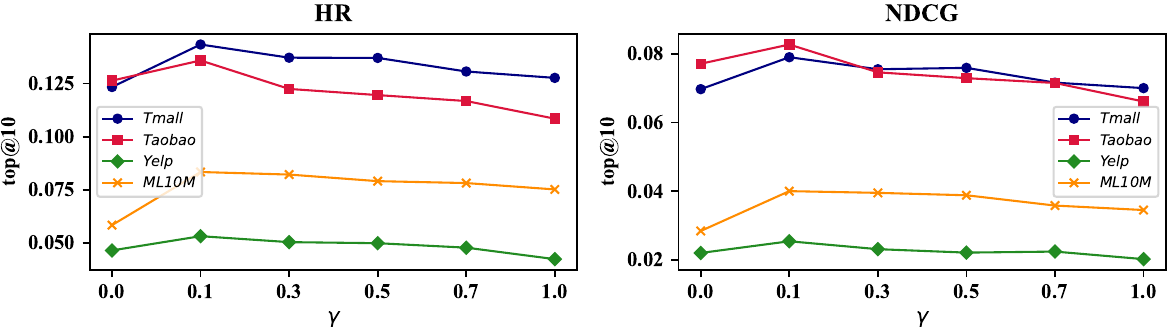}
    \caption{Analysis of the hyperparameter $\gamma$ (x-axis represents the values of $\gamma$, and y-axis represents the model's performance).}
    \label{fig:gamma}
    \Description{}
\end{figure}

\textit{\textbf{Hyperparameter $\boldsymbol{\gamma}$}}. 
$\boldsymbol{\gamma}$ is a parameter used to control the degree of the KL divergence constraint in the VAE module. The KL divergence loss primarily forces the encoder to encode information into latent representations that are close to a standard normal distribution, thereby enhancing the model's generalization ability. To investigate its impact on overall performance, we analyze the variations in performance when $\boldsymbol{\gamma}$ takes different values. The results are shown in Fig.~\ref{fig:gamma}.

From the experimental results, it can be observed that across all four datasets, the model achieves optimal performance on both HR and NDCG metrics when $\boldsymbol{\gamma}$ is set to 0.1. Subsequently, the model's performance declines as $\boldsymbol{\gamma}$ increases. The superior performance of the model when $\boldsymbol{\gamma}$ = 0.1 compared to $\boldsymbol{\gamma}$ = 0.0 confirms that the KL divergence loss contributes positively to model training. However, when $\boldsymbol{\gamma}$ exceeds 0.1, the model's performance continues to deteriorate. This decline is attributed to the excessively large KL divergence weight, which may lead the model to learn too simplistic or overly constrained latent representations, making it difficult to capture the complex features and structures in the data, thereby reducing the overall performance.

\subsection{Study of Cold-start Problem}

The cold-start problem is a common challenge in recommendation systems. It arises when the system encounters new users or new items, and due to a lack of sufficient historical data or information, traditional recommendation algorithms often fail to provide accurate personalized recommendations for these new entities. Multi-behavior recommendation leverages users' diverse behavioral data to enhance recommendation capabilities, which can partially alleviate the cold-start problem. In this section, we discuss our model's performance in addressing the cold-start issue.

\begin{figure}[ht]
    \centering
    \includegraphics[width=0.8\linewidth]{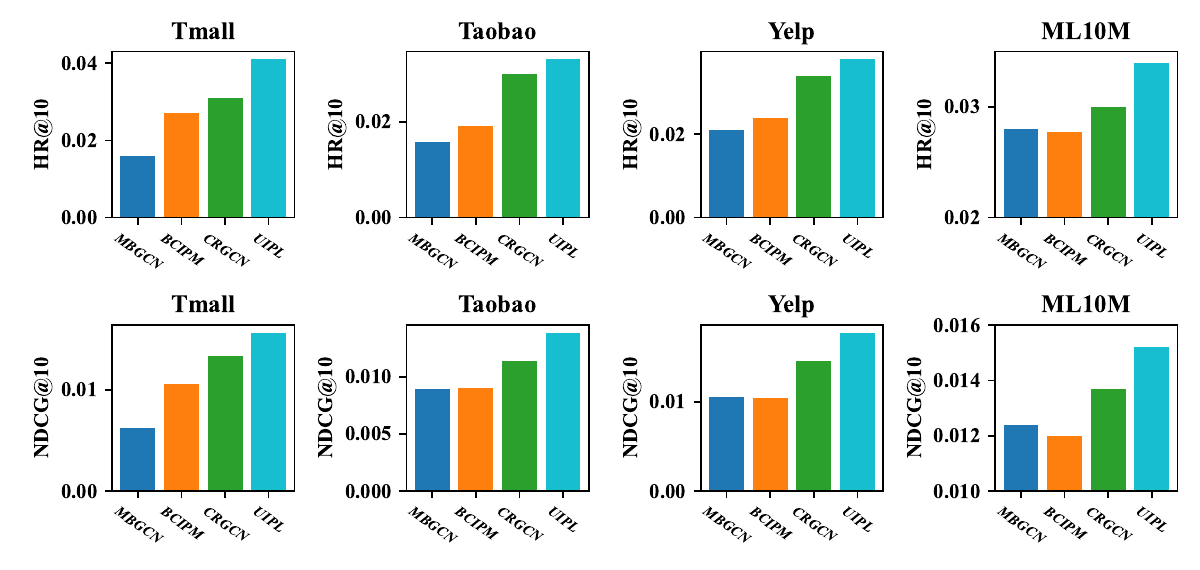}
    \caption{Performance comparisons among MBGCN, CRGCN, BCIPM and UIPL for cold-start users.}
    \label{fig:cold_start}
    \Description{}
\end{figure}

We compare our model with three representative methods: MBGCN, CRGCN, and BCIPM. MBGCN incorporates an item-item relation computation module to mitigate the cold-start problem by calculating the similarity between a specific item and items previously interacted with by the user. CRGCN refines user preferences sequentially, and when no interactions exist in the target behavior, it can leverage coarse-grained information for predictions. BCIPM serves as the strongest baseline method. To conduct this study, we randomly select 1,000 users from the test set as a new cold-start user test set (denoted as "$test_{c}$") and remove all records of their target behaviors from the training set. Additionally, to ensure the system has no prior knowledge about users' preferences for test items, we remove all user-item interaction pairs appearing in "$test_{c}$" set from the auxiliary behavioral interaction records. We then train the model using the remaining training data and evaluate its performance on the "$test_{c}$" dataset. The specific experimental results are shown in Fig~\ref{fig:cold_start}.

The experimental results show that UIPL significantly outperforms the other three methods across all four datasets, highlighting the effectiveness of its design. UIPL leverages IRM to capture user invariant preferences, which brings two key benefits: (1) avoiding noise introduced by auxiliary behaviors and (2) enabling recommendations based on consistent user invariant preferences even in the absence of target behavior interactions. When target behavior data is sparse or absent, UIPL loses the ability to capture preferences specific to the target behavior but retains invariant user preferences learned from auxiliary behaviors, allowing for relatively accurate predictions. Consequently, UIPL achieves superior performance compared to the other methods. 

In contrast, MBGCN introduces an item-item relevance computation module to address the cold-start problem but employs a simplistic fusion strategy for multi-behavior data modeling. This approach introduces substantial noise from auxiliary behaviors, which adversely affects target behavior predictions, particularly in scenarios with limited target behavior interactions. In such cases, the noise dominates, leading to subpar performance. CRGCN performs better than MBGCN by incorporating sequential modeling, which refines user preferences iteratively. Each subsequent behavior adjusts and corrects the preceding ones, effectively reducing noise compared to MBGCN. While BCIPM is the strongest baseline overall, it performs poorly in cold-start scenarios due to its primary focus on denoising. By using auxiliary behaviors solely to enhance training without adequately incorporating relevant information from these behaviors, BCIPM fails to effectively address the cold-start problem.

These findings suggest that while multi-behavior methods can alleviate the cold-start issue to some extent, their effectiveness heavily depends on the modeling strategies employed.

\section{Conclusion}

In this work, we conducted an in-depth analysis of the noise introduction problem in multi-behavior data modeling and proposed a solution. Our approach decomposes user preferences for each behavior into invariant preferences and behavior-specific preferences. By leveraging a variational autoencoder to reconstruct user invariant preferences and applying an invariant risk minimization paradigm for reconstruction constraints, we successfully achieved the separation of invariant and behavior-specific preferences. Extensive experiments conducted on four datasets validated the effectiveness of the proposed model.

In our future work, we plan to delve deeper into the balance between invariant preferences and target-behavior-specific preferences, an aspect not fully addressed in this study but deemed crucial. As our research progresses, identifying the optimal balance between these two types of preferences will be critical to improving the performance of recommendation systems. Moreover, despite the stark deviation of the method employed in this paper from conventional multi-behavior recommendation approaches, we still face challenges due to the entanglement of user preference representations, impeding the system's ability to discern users' fine-grained preferences effectively. Consequently, existing methods exhibit certain limitations in enhancing recommendation accuracy and interpretability. To tackle this challenge, we will focus on decoupling invariant preferences to extract more granular user preference information. This endeavor not only aids in better capturing users' true intentions but also significantly boosts the interpretability of recommendation systems, fostering user trust and acceptance of the recommended outcomes.

\bibliographystyle{ACM-Reference-Format}
\bibliography{sample-base}

\end{document}